\documentclass[aps,pra,floatfix,twocolumn,showpacs]{revtex4}

\newcommand{\szn}{{\langle s_{z_n} \rangle}}
\newcommand{\szp}{{\langle s_{z_p} \rangle}}
\newcommand{\sz} {{\langle s_{z}   \rangle}^o}
\newcommand{\jzp}{{\langle j_{z_p} \rangle}}
\newcommand{\lzp}{{\langle l_{z_p} \rangle}}
\newcommand{\lzn}{{\langle l_{z_n} \rangle}}
\newcommand{\rr} {{r_{_{12}}}}

\usepackage{amssymb}
\usepackage{amsfonts}
\usepackage{amsthm}
\usepackage{latexsym,amsmath}
\usepackage[dvips]{graphicx}
\usepackage{epsfig}
\usepackage{graphicx}
\usepackage{dcolumn}

\begin{document}
\title{Dependence of nuclear magnetic moments on quark masses and
limits on temporal variation of fundamental constants from atomic
 clock experiments}
\author{V.V. Flambaum, A.F. Tedesco}
\affiliation{
 School of Physics, The University of New South Wales, Sydney NSW
2052, Australia
}
\date{\today}
\begin{abstract}
 We calculate the dependence of the nuclear magnetic moments on the quark
 masses including the spin-spin interaction effects and  obtain limits on
 the variation of the fine structure constant $\alpha$ and $(m_q/\Lambda_{QCD})$ using recent atomic clock experiments examining hyperfine transitions in H, Rb, Cs, Yb$^+$ and Hg$^+$ and the optical transition in H, Hg$^+$ and Yb$^+$ . 

\end{abstract}

\maketitle

\section{Introduction}
  Theories unifying gravity with other interactions suggest a possibility
of temporal and spatial variation of the fundamental constants of nature
(see e.g. review \cite{Uzan} where the theoretical models and
results of measurements are presented). There are hints for the variation
of the fundamental constants in Big Bang nucleosynthesis, quasar absorption
spectra and Oklo natural nuclear reactor data. However, a majority
of publications report only limits on  possible variations. For example,
comparison of different atomic clocks gives limits on present time
 variation of the fundamental constants.

 A large fraction of the publications discuss the variation
 of the fine structure constant $\alpha = e^2/\hbar c$.
 The hypothetical unification of all interactions implies that a variation
 in $\alpha$ should be accompanied by a variation of the strong interaction
strength and the fundamental masses. For example, the grand unification model
 discussed in Ref. \cite{Langacker 2002} predicts the quantum chromodynamics
 (QCD) scale $\Lambda_{QCD}$ (defined as the position of the Landau pole in
 the logarithm for the running strong coupling constant) is modified as $
{\delta\Lambda_{QCD}}/{\Lambda_{QCD}} \approx 34 \;{\delta\alpha}/{\alpha}$.
The variation of quark and electron masses in this model is given by
${\delta m}/{m} \sim 70 \; {\delta\alpha}/{\alpha} $,
giving an estimate of the variation for the dimensionless ratio
\begin{equation}
\label{eq:alpha}
\frac{\delta (m_q/\Lambda_{QCD})}{(m_q/\Lambda_{QCD})} \sim 35\frac{\delta\alpha}{\alpha}
\end{equation}

The coefficient here is model dependent but large values are generic for
 grand unification models in which modifications come from high energy scales;
 they appear because the running strong-coupling constant and Higgs constants
 (related to mass) run faster than $\alpha$. If these models are correct,
 the variation in quark masses and the strong interaction scale may be easier
 to detect than a variation in $\alpha$.

One can only measure the variation of dimensionless quantities. We want to
 extract from the measurements the variation of the dimensionless ratio
 $m_q/\Lambda_{QCD}$ where $m_q$ is the quark mass (with the dependence on
 the renormalization point removed). A number of limits on the variation of
 $m_q/\Lambda_{QCD}$ have been obtained recently from consideration of Big
 Band nucleosynthesis, quasar absorption spectra and the Oklo natural nuclear
 reactor, which was active about 1.8 billion years ago
 \cite{Flambaum Shuryak 2002,
 Dmitriev Flambaum 2003, Flambaum Shuryak 2003}.

Karshenboim \cite{Karshenboim 2000} has pointed out that measurements of ratios
 of hyperfine structure intervals in different atoms are sensitive to
 variations in nuclear magnetic moments. However, the magnetic moments
are not the fundamental parameters and can not be directly compared with any
theory of the variations. Atomic and nuclear calculations are needed
for the interpretation of the measurements. Below, we calculate the dependence
 of nuclear magnetic moments on $m_q/\Lambda_{QCD}$ by building on recent work
 and incorporating the effect of the spin-spin interaction  between nucleons.
 We obtain limits
 on the variation of $m_q/\Lambda_{QCD}$  from recent experiments that
 have measured the time dependence of
 the ratios of the hyperfine structure intervals of $^{133}$Cs and
 $^{87}$Rb \cite{Bize 2005},
 $^{133}$Cs and $^{1}$H \cite{Cs},
 $^{171}$Yb$^+$ and $^{133}$Cs \cite{Peik 2004}, $^{199}$Hg$^+$ and $^{1}$H \cite{Prestage 1995},
  the ratio of the optical frequency
 in $^{199}$Hg$^+$ to the hyperfine frequency of $^{133}$Cs \cite{Bize 2003},
  the ratio of the optical frequency
 in $^{1}$H to the hyperfine frequency of $^{133}$Cs \cite{Fischer 2004},
and  the ratio of the optical frequency in $^{171}$Yb$^+$ to the hyperfine
 frequency of $^{133}$Cs \cite{Peik 2005}.
It has been suggested  in Ref.\cite{Flambaum 2006} that the
effects of the fundamental constants variation may be enhanced
2-3 orders of magnitude in diatomic molecules like LaS, LaO, LuS, LuO.
Therefore, we also present the results for $^{139}$La.

During the calculations, we shall assume (for notational convenience)
 that the strong interaction scale $\Lambda_{QCD}$ does not vary and so
 we shall speak about the variation of masses (this means that we measure
 masses in units of $\Lambda_{QCD}$). We shall restore the explicit
 appearance of $\Lambda_{QCD}$ in the final answers.

The hyperfine structure constant can be presented as
\begin{equation}
A = const \times (\frac{m_e e^4}{\hbar^2})\, [\alpha^2 F_{rel}(Z\alpha)]\, (\mu\frac{m_e}{m_p})
\end{equation}
The factor in the first set of brackets is an atomic unit of energy.
 The second ``electromagnetic'' set of brackets determines the dependence
 on $\alpha$ and includes the relativistic correction factor (Casimir factor)
 $F_{rel}$ . The last set of brackets contains the dimensionless nuclear
 magnetic moment $\mu$ (that is, the nuclear magnetic moment
 $M = \mu [e \hbar / 2 m_p c]\:)$ and the electron and proton masses
 $m_e$ and $m_p$. We might also have included a small correction due to
 the finite nuclear size but its contribution is insignificant.

The ratio of two hyperfine structure constants for different atoms will
 cancel out some factors such as atomic unit of energy and $m_e/m_p$,
 and any time dependence falls
 on two values: the ratio of the factors $F_{rel}$
 (which depends on $\alpha$) and the ratio of the nuclear magnetic moments
 (which depends on $m_q/\Lambda_{QCD}$). 

For the $F_{rel}$ component, variation in $\alpha$ leads to the following
 variation of $F_{rel}$ \cite{Prestage 1995}
\begin{equation}
\frac{\delta F_{rel}}{F_{rel}} = K_{rel} \frac{\delta \alpha}{\alpha}
\end{equation}
and one can use the $s-$wave electron approximation for $F_{rel}$
 to get
\begin{equation}\label{K}
 K_{rel} = \frac{(Z\alpha)^2(12\gamma^2-1)}
{\gamma^2(4\gamma^2-1)}
\end{equation}
 where $ \gamma = \sqrt{1-{(Z\alpha)}^2}$. 
However, numerical many-body calculations \cite{Dzuba 1999} give more
 accurate results, with a slightly higher value of $K_{rel}$ than that
 given by this formula. A comparison is shown in Table \ref{tab:aa}.

%
%
%

The other component contributing to the ratio of two hyperfine structure
 constants is the nuclear magnetic moment $\mu$. Theoretical values for
 $\mu$ in the valence shell model are based on the unpaired valence nucleon
 and are given by Schmidt values
\begin{equation}\label{mu}
\mu = \left \{ \begin{array}{l l}
\frac{1}{2}[g_s + (2j-1)g_l] & \mbox{for $j=l+\frac{1}{2}$} \vspace{2mm}\\
\frac{j}{2(j+1)}\left [-g_s+(2j+3)g_l \right ] &
 \mbox{for $j=l-\frac{1}{2}$} \end{array} \right.
\end{equation}

The orbital gyromagnetic factors are $g_l = 1$ for a valence proton and
 $g_l = 0$ for a valence neutron. The spin gyromagnetic factors are $g_s \,
 (= g_p) = 5.586$ for protons and $g_s \, (= g_n) = -3.826$ for neutrons.
 These $g$-factors depend on $m_q/\Lambda_{QCD}$ and previous work exploring
 this dependence \cite{ Flambaum 2003,Flambaum 2004}  is summarized below.
 We then use these results to consider the more realistic situation of
 $\mu$ having both a valence nucleon contribution and a non-valence nucleon
 contribution due to the spin-spin interaction. 

\section{Variation of magnetic moment with variation in quark mass}

\subsection{Variation in $\mu$ using  valence model magnetic moment}

As a preliminary to our results and as a comparison for evaluating the effects
 of our calculations, we include the results of work previously done in this
 area \cite{ Flambaum 2003,Flambaum  2004}. This work was  essential to our
 results as the authors calculated the variation in the neutron and proton
 magnetic moments ($\mu_n$ and $\mu_p$) with  the variation in
 $m_q/\Lambda_{QCD}$ using chiral perturbation theory.

As mentioned above, the $g$-factors depend on $m_q/\Lambda_{QCD}$.
 The light quark mass $m_q = (m_u + m_d)/2 \approx 5$MeV and in the chiral
 limit $m_u=m_d=0$, the nucleon magnetic moment remains finite. Thus one might
 assume that corrections to the spin $g$-factors $g_p$ and $g_n$ are small.
 However, the quark mass contribution is enhanced by $\pi$-meson loop
 corrections to the nuclear magnetic moments, which are proportional to
 $\pi$-meson mass $m_{\pi} \sim  \sqrt{m_q \Lambda_{QCD}}$.
 Since $m_\pi$ = 140 MeV, the contribution can be significant.

Full details of these calculations are given in Ref \cite{ Flambaum 2003,
Flambaum  2004}. They give the following results, which relate variations
 in $\mu_n$ and $\mu_p$ with variations in light and strange quark masses
 ($m_q$ and $m_s$):
\begin{eqnarray}\label{mq}
\label{eq: mupq} \frac{\delta\mu_p}{\mu_p} &= & - \, 0.087 \:\, \frac{\delta m_q}{m_q}\\
\label{eq: mups} \frac{\delta\mu_p}{\mu_p} &= & - \, 0.013 \:\,  \frac{\delta m_s}{m_s}\\
\label{eq: munq} \frac{\delta\mu_n}{\mu_n} &= & - \, 0.118 \:\, \frac{\delta m_q}{m_q}\\
\label{eq: muns} \frac{\delta\mu_n}{\mu_n} &= & + \, 0.0013 \frac{\delta m_s}{m_s}
\end{eqnarray} 

Using these relations and the valence model approximations for $\mu$,
 we can obtain expressions of the form
\begin{equation}
\frac{\delta\mu}{\mu} =  \kappa_q \frac{\delta m_q}{m_q} + \kappa_s \frac{\delta m_s}{m_s}
\end{equation}
Hence for nuclei with even Z and a valence neutron
\[ \frac{\delta\mu}{\mu} = \frac{\delta g_n}{g_n } = -0.118
 \frac{\delta m_q}{m_q } +0.0013 \frac{\delta m_s}{m_s}.\]


\begin{table*}
\caption{Variational factor $K_{rel}$ and $\kappa$ values for various atoms obtained using
 simple valence shell model (method A) as used in
 equation (\ref{eqn:V}). The first
row is given by equation (\ref{K}). The second row presents the results
of the more accurate many-body calculations (see ref. \cite{Dzuba 1999}). The numerical
results marked by $^*$ are obtained by an extrapolation from other atoms.}
\label{tab:aa}
\vspace{2mm}
\centering
\begin{tabular}{@{\extracolsep{5mm} } c|ccccccccc }
\hline
\hline
\\[-2mm]
Atom  & $^1$H & $^2$H & $^3$He & $^{87}$Rb & $^{111}$Cd & $^{129}$Xe & $^{133}$Cs & $^{171}$Yb & $^{199}$Hg \\
\\[-3mm] \hline \\[-2.3mm]
$K_{rel}$ (analytical) & 0 & 0 & 0 & 0.29 & 0.53  & 0.71 & 0.74 & 1.42 & 2.18 \\
\\[-3mm] \hline \\[-2.3mm]
$K_{rel}$ (numerical) & 0 & 0 & 0 & 0.34 & $0.6^*$  & $0.8^*$ & 0.83 & $1.5^*$ & 2.28 \\
\\[-3mm] \hline \\[-2.3mm]
$\kappa_q $ & -0.087 & -0.020 & -0.118 & -0.064 & -0.118 & -0.118 & 0.110 & -0.118 & -0.118 \\
\\[-3mm] \hline \\[-2.3mm]
$\kappa_s $ & -0.013 & -0.044 & 0.0013 & -0.010 & 0.0013 & 0.0013 & 0.016 & 0.0013 & 0.0013 \\
\hline
\end{tabular}
\end{table*}

 For valence protons, the orbital gyromagnetic factor $g_l$ also has an
 effect. Thus for $^{133}$Cs with its valence proton and $j=l-\frac{1}{2}$,
\[ \frac{\delta\mu}{\mu}  = 0.110 \frac{\delta m_q}{m_q }
 +0.016 \frac{\delta m_s}{m_s}\] 
while for $^{87}$Rb with its valence proton but $j=l+\frac{1}{2}$ ,
\[ \frac{\delta\mu}{\mu}  = - 0.64 \frac{\delta m_q}{m_q } - 0.010
 \frac{\delta m_s}{m_s}\] 

These results can be presented using the ratio of the hyperfine constant $A$ to the atomic unit of energy $E = m_ee^4/\hbar^2$ by defining the parameter $V$ through the relation
\begin{equation} \label{eqn:V def}
\frac{\delta V}{V} = \frac{\delta (A/E)}{A/E}.
\end{equation}

The values for $\kappa_q$ and $\kappa_s$ in the results for $\delta\mu /\mu$ can then be combined with the corresponding values of $K_{rel}$ in Table \ref{tab:aa} to give results of the form:
\begin{equation} \label{eqn:V} 
V(M)  =  \alpha^{2+K_{rel}} \left(\frac{m_q}{\Lambda_{QCD}}\right) ^{\kappa_q} \left(\frac{m_s}{\Lambda_{QCD}}\right)^{\kappa_s}  \frac{m_e}{m_p} 
\end{equation}


The calculated values of $K_{rel}$ and $\kappa$ to be used in the expression for V(M) for various atoms are summarized in Table \ref{tab:aa}.
The factor $m_e/m_p$ will cancel out when a ratio of hyperfine transitions is used. It will, however, survive in a comparison with optical and molecular transitions.

\subsection{Variation in $\mu$ incorporating the effect of non-valence nucleons by using the experimental magnetic moment}

The results of the previous section were used to calculate the variation of $\mu$ with $m_q/\Lambda_{QCD}$ based on the single particle approximation for $\mu$ (one valence nucleon) within the shell model. That is, it was assumed that the dimensionless nuclear magnetic moment is given by
$\mu =  g_p \, \sz + \langle l_{z} \rangle^{0}$ for a valence proton, and
$\mu = g_n \, \sz$ for a valence neutron.
Here, $g_p$ and $g_n$ are the spin gyromagnetic factors for free protons and neutrons respectively,  $\langle l_{z} \rangle^{0}=j_z-\sz$, and $\sz$ is the
 spin expectation value of the single valence nucleon in shell model:
\begin{equation}
\sz = \left \{ \begin{array}{l l}
\frac{1}{2} & \mbox{for $j=l+\frac{1}{2}$}\\
-\frac{j}{2(j+1)} & \mbox{for $j=l-\frac{1}{2}$} \end{array} \right.
\end{equation}

However, it is well known that this theoretical value is only an estimate of
 $\mu$ and the magnetic moment of the valence nucleon tends to be offset by
 a  contribution from the  core nucleons.
An empirical rule is that the spin contribution of a valence
nucleon should be reduced by 40\% to obtain a reasonable value
for the nuclear magnetic moment. This reduction may be explained
by the contribution of core nucleons, which should be negative since
proton and neutron magnetic moments are large and have opposite signs. 

For example, a valence proton polarizes, by the spin-spin interaction,
core neutrons and these core neutrons give a negative contribution
to the nuclear magnetic moment (polarization of the core protons  by
the valence proton is not important). 
 We can estimate this offset by
 considering  contributions to $\mu$ from the valence and core
 nucleons. This means we have both neutron and proton spin contributions
 to $\mu$:
\begin{equation}
\label{eq:mu1} \mu = g_n \, \szn + g_p \, \szp + \lzp
\end{equation}
We neglected here a small contribution of the exchange currents
into the magnetic moment.

 We want to evaluate the corrections to the valence
model results using the very accurate experimental values of
nuclear magnetic moments. Since there are three unknown parameters
($\szn , \szp , \lzp $) and only one experimental value
(the total magnetic moment) available,
to perform an estimate we need to make approximations.

However, as we will show below, the result is not sensitive to particular
approximations if we can reproduce the experimental magnetic moment exactly.
Indeed, using the equations above, we can present the variation
 of the magnetic moment in the following form
\begin{equation}
\label{eq:delta mu1} \delta \mu- \delta \mu^o =
[0.45(\szn -\szn^o)  -0.56(\szp -\szp^o)]\frac{\delta m_q}{m_q }
\end{equation}
Here  $\delta \mu^o$ is the valence model value.
For brevity we here assume that $\frac{\delta m_q}{m_q }=
 \frac{\delta m_s}{m_s}$ (the coefficient before $\frac{\delta m_s}{m_s}$
is small anyway)). 

Let us start with the case of a valence proton.
 The simplest assumption is that the spin-spin interaction
transfers part of the proton spin to the  core neutron spin, i.e.
$(\szn -\szn ^o)=-(\szp -\szp ^o)$ and $\lzp=\lzp ^o$. Then we can solve
the equation for the magnetic moment and obtain the deviation
from the valence model value
\begin{equation}
\label{eq:delta 1} \delta \mu- \delta \mu^o =
-0.11 (\mu-  \mu^o) \frac{\delta m_q}{m_q }
\end{equation}
To test the stability of the result, we can try different ``extreme'' assumptions.
For example, if the angular momentum exchange occurs exclusively between the proton
spin and proton orbital angular momentum, then $\lzp-\lzp ^o=-(\szp -\szp ^o)$,
$\szn =\szn ^o$. In this case
\begin{equation}
\label{eq:delta 2} \delta \mu- \delta \mu^o =
-0.12 (\mu-  \mu^o) \frac{\delta m_q}{m_q }
\end{equation}
Finally, we can try an unreasonable assumption that
the exchange happens between the proton
spin and neutron orbital angular momentum: $\lzn-\lzn ^o=-(\szp -\szp ^o)$,
$\szn =\szn ^o$, $\lzp=\lzp ^o$. Then
\begin{equation}
\label{eq:delta 3} \delta \mu- \delta \mu^o =
-0.10 (\mu-  \mu^o) \frac{\delta m_q}{m_q }
\end{equation}
We see that the results are very stable, the difference in the correction
to the valence model is about 10 \% only. The results for a valence
neutron are similar. The coefficients are -0.11
 for $(\szn -\szn ^o)=-(\szp -\szp ^o)$,
 -0.12   for $\lzn-\lzn ^o=-(\szn -\szn ^o)$, and
-0.09 for $\lzp-\lzp ^o=-(\szn -\szn ^o)$.

 To present the final results, we will use a physical approximation which gives
results somewhere in between the ``extreme'' assumptions.
The nuclear magnetic moments are reproduced with a reasonable
accuracy by the RPA calculations.    
In the RPA approximation there are two separate conservation
laws for the total proton $j_p$ and total neutron $j_n$ angular momenta
 (see e.g. \cite{Dmitriev 1983}).
We also assume that total orbital angular momentum $\lzn +\lzp$
and total spin  $\szn +\szp$ are conserved
(this assumption corresponds to  neglect of
 the spin-orbit interaction).
 We repeat again that we only need these approximations to obtain
 specific numbers
which are in between ``extreme'' model values. Then 
 we can write
\begin{eqnarray}
\label{eq:sz} \sz & = & \szp + \szn \\
\label{eq:jzp} \jzp & = & \lzp + \szp
\end{eqnarray}
where $\jzp =I$ for a valence proton and $\jzp = 0 $ for a valence neutron.
 Using equations (\ref{eq:jzp}) and (\ref{eq:sz}) to eliminate
$\langle l_{z_p} \rangle$ and  $\langle s_{z_p} \rangle$
 in equation (\ref{eq:mu1}) we get
\begin{equation}
\label{eq:mu2} \mu  = g_n \szn + (g_p - 1)\szp + \jzp
\end{equation}
\begin{equation}
\label{eq:szn} \szn = \frac{\mu - \jzp - (g_p - 1)\sz}{g_n + 1 -g_p}
\end{equation}
\begin{equation}
\label{eq:szp} \szp = \sz - \szn
\end{equation}
We thus have taken into account both the proton and neutron
 contributions to the nuclear magnetic moment and can more accurately
 estimate how a variation in quark mass relates to a variation in $\mu$.
 From equation (\ref{eq:mu1}) we see immediately that
\begin{equation}
\delta\mu = \delta g_n \szn + \delta g_p \szp
\end{equation}
and thus
\begin{equation}
\label{eq:K}
\frac{\delta\mu}{\mu} = \frac{\delta g_n}{g_n} K_n + \frac{\delta g_p}{g_p} K_p
\end{equation}
where 
\[
K_n  =  \left \{ \frac{g_n \szn}{\mu} \right \} \mbox{\: \: and \: \:}
K_p  =  \left \{ \frac{g_p \szp}{\mu} \right \}\]

From the definition of the $g$-factor for free protons and free neutrons,
 we know $\delta g_n / g_n = \delta\mu_n / \mu_n$ and $\delta g_p /
 g_p = \delta\mu_p / \mu_p$. We can now use equations
 (\ref{eq: mupq} - \ref{eq: muns}) to explicitly relate the variation in
 $\mu$ to the variation in quark masses. Thus
\begin{equation}
\label{eq:mu5}
\frac{\delta\mu}{\mu} =  \kappa_q \frac{\delta m_q}{m_q} + \kappa_s \frac{\delta m_s}{m_s}
\end{equation}
where clearly
\[ \kappa_q =  -0.118 K_n -0.087 K_p  \] 
\[\kappa_s =  0.0013 K_n -0.013 K_p  \]

We are now in a position to evaluate the coefficients in specific cases.
 For ${}^{133}$Cs, $I^{\pi}= {7/2}^{+}$ and $\mu = 2.5820$ and it has
 a valence proton. Thus $\jzp = \frac{7}{2} = l-\frac{1}{2}$ and
 $\sz = - \frac{7}{18}$. Therefore equations (\ref{eq:szn}) and
 (\ref{eq:szp}) immediately give us
\[ \szn = -0.103 \]
\[ \szp = -0.286 \]
and thus
\[ \frac{\delta\mu}{\mu} = 0.152 \frac{\delta g_n}{g_n} -0.619 \frac{\delta g_p}{g_p}\]
giving
\[ \frac{\delta\mu}{\mu} = 0.0358 \frac{\delta m_q}{m_q } + 0.00824 \frac{\delta m_s}{m_s}.\]
The dependence on the strange quark mass is relatively weak and it is
 convenient to assume that the relative variation of the strange quark mass
 is the same as the relative variation in the light quark masses
 (this assumption is motivated by the Higgs mechanism of mass generation).
 We restore the explicit notation for the strong-coupling constant and
 conclude
\begin{equation}
\frac{\delta\mu}{\mu} = 0.0441 \frac{\delta
 (m_q/\Lambda_{QCD})}{(m_q/\Lambda_{QCD})} \mbox{\; \; for \; $^{133}$Cs.}
\end{equation}
\vspace{1mm}

\begin{table*}
\caption{Variation in $\mu$ incorporating the effect of non-valence nucleons in various atoms. $\sz$ is the spin expectation value, $\szn$ and $\szp$ are (either, depending on nucleus) the valence and non-valance nucleon contributions to this spin, $K_n$ and $K_p$ are defined in equation \ref{eq:K}, and $\kappa_q$ and $\kappa_s$ are defined in equation \ref{eq:mu5}.}
\label{tab:b}
\vspace{2mm}
\centering

\begin{tabular}{@{\extracolsep{6mm}}c|ccccccccc}
\hline
\hline
\\[-2.5mm]
Atom & $^1$H & $^2$H & $^{3}$He & $^{87}$Rb & $^{111}$Cd & $^{129}$Xe & $^{133}$Cs & $^{171}$Yb & $^{199}$Hg \\    
\\[-3mm] \hline \\[-2.5mm]
$\sz$ & 0 & 1 & 0.5 & 0.5 & 0.5 & 0.5 & -0.389 & -0.167 & -0.167 \\
\\[-3mm] \hline \\[-2.5mm]
$\szn$ & 0 & 0.5 & 0.5 & 0.124 & 0.343 & 0.365 & -0.103 & -0.150 & -0.151  \\
\\[-3mm] \hline \\[-2.5mm]
$\szp$  &  0.5 & 0.5 & 0 &   0.376 & 0.157 & 0.135  & -0.286 & -0.017 & -0.016  \\
\\[-3mm] \hline \\[-2.5mm]
$K_n$  & --  & -- & -- &  -0.172 &  2.21  &  1.80  &  0.152  &  1.16  &  1.14   \\
\\[-3mm] \hline \\[-2.5mm]
$K_p$ & --  & --  &  --  & 0.764 &  -1.47 & -0.969 &  -0.619  & -0.194 &  -0.173  \\
\\[-3mm] \hline \\[-2.5mm]
$\kappa_q$ & -0.087 & -0.020 & -0.118 & -0.046 & -0.133 & -0.128 &  0.036 & -0.120 &  -0.120  \\
\\[-3mm] \hline \\[-2.5mm]
$\kappa_s$ & -0.013 & -0.044 & 0.0013 & -0.010 & 0.022 & 0.015 & 0.008 & 0.004 & 0.004 \\
\\[-3mm] \hline \\[-2.5mm]
$\kappa_q +\kappa_s$ & -0.100  & -0.064 & -0.117 & -0.056 & -0.111 & -0.113 & 0.044 & -0.116 & -0.116 \\
\hline
\end{tabular}
\end{table*}


Values for ${}^{87}$Rb, ${}^{199}$Hg, ${}^{171}$Yb, ${}^{111}$Cd and ${}^{129}$Xe can be similarly calculated and are presented in Table \ref{tab:b} 

For $^2$H and $^3$He, the magnetic moments are pretty close
to naive values, therefore we have not tried to improve the results.

These results can be summarized in Table \ref{tab:b}. A comparison with the
 earlier results (using the valence model method) for the total
 variational relation ($\kappa_q + \kappa_s$) is shown later in
 Table \ref{tab:c}.
We see that $\delta\mu/\mu$ in the nuclei with a valence proton is very
sensitive to the core polarization effects. However, there is no
such sensitivity in the nuclei with a valence neutron. To explain this
conclusion one should note that a neutron does not give an orbital
contribution to the nuclear magnetic moment.
 The orbital contribution of core protons is relatively small.
 As a result, the core
polarization effect changes $\delta\mu$ and $\mu$ in a similar
way, i.e. it practically does not change their ratio. In the nuclei with
a valence neutron
\begin{equation}
\frac{\delta\mu}{\mu} = -0.117 \frac{\szn -1.25 \szp}{\szn -1.20 \szp}
\frac{\delta m_q}{m_q}
\end{equation}

\subsection{Effect of variation of the the spin-spin interaction}
In previous subsection we have not taken into account
 that $\szp$  and $\szn$ may depend on  the quark mass.
However, this dependence appears since the spin-spin interaction
depends on the quark masses. Below we want to perform a rough estimate of
this effect using the one-boson-exchange model of the strong interaction.

Consider, for example, a nucleus with a valence proton.
As above we assume that the total  spin of nucleons is conserved and so
\begin{eqnarray}
\sz & = & \szp  + \szn
\end{eqnarray}
where $\sz$ is again the valence model value.
It is convenient to use the following notations:
\begin{eqnarray}
\szn & = & b \sz  \\
\szp & = & (1-b) \sz  
\end{eqnarray}
where b is a coefficient determined by the spin-spin interaction. We need to determine the dependence of b on $m_q/\Lambda_{QCD}$ to complete our calculations. It can be estimated using perturbation theory as
\begin{eqnarray}
b & \sim & \left ( \frac{ \langle o | V_{ss} | k \rangle }{E_o - E_k} \right )^2
\end{eqnarray}
where \[ V_{ss} = V ( | \mathbf{r}_1 - \mathbf{r}_2 | ) \mathbf{S}_1 \cdot \mathbf{S}_2\]
and $E_o - E_k$ is the spin-orbit splitting (see e.g.\cite{Dmitriev 1983}).
 As $m_q \rightarrow 0$, $E_o - E_k$ remains finite and so it can only
 have a weak dependence on $m_q/ \Lambda_{QCD}$. The major dependence comes
 from the $\pi$-meson mass which vanishes
in the chiral limit  $m_q \rightarrow 0$.
 According to review \cite{spin}
 the $\pi$-meson exchange gives about 1/3 of the spin-spin interaction.
The other most significant contribution is given by the  $\rho$-meson exchange.
We neglect other meson contributions and assume that the remaining 2/3
 of the spin-spin
interaction is given  by the  $\rho$-meson. The result is not very
 sensitive to this assumption since
 the $\rho-$meson and other vector mesons
have approximately the same and rather weak sensitivity to a variation
 in $m_q$. According to Ref. \cite{roberts}
\[ \frac{\delta m_{\rho}}{m_{\rho}} = 0.021 \frac{\delta m_q}{m_q }\] 
whereas for the $\pi-$meson
\[ m_{\pi} \sim  \sqrt{m_q \Lambda_{QCD}} \]
so we have
\[ \frac{\delta m_{\pi}}{m_{\pi}} = \frac{1}{2} \frac{\delta m_q}{m_q }.\]

The dominating contribution is therefore given by $\pi$-meson exchange.
  The exchange contribution of $\pi^{\pm}$ is small due to the small
 overlap between $\psi_p(r)$ and $\psi_n(r)$ and is not important.
 The main contribution is when a neutron is excited through $\pi^0$ exchange
 into a spin-orbit doublet, $j=l+\frac{1}{2}$ to $j=l-\frac{1}{2}$.
  The strongest dependence thus originates from the $\pi^0$ pion mass.

The momentum space representation of the nucleon-nucleon interaction due to a $\pi$-meson is
\[ V_\pi(q) = g_\pi^2(\vec{\tau}_1 \cdot \vec{\tau}_2)(\vec{\sigma}_1 \cdot \mathbf{q})(\vec{\sigma}_2 \cdot \mathbf{q}) \frac{1}{m_\pi^2 + q^2} \]
where q is the momentum transfer $\mathbf{q} = \mathbf{p}_1 - \mathbf{p}_2$, $\vec{\tau}$ is the isotopic spin and $\vec{ \sigma }$ is the Pauli spin matrix.

We separate this into tensor and scalar parts 
\begin{eqnarray}
\nonumber (\vec{\sigma}_1 \cdot \mathbf{q})(\vec{\sigma}_2 \cdot \mathbf{q}) & = \left \{  (\vec{\sigma}_1 \cdot \mathbf{q})(\vec{\sigma}_2 \cdot \mathbf{q}) - \frac{1}{3} \vec{\sigma}_1 \cdot \vec{\sigma}_2 q^2 \right \} \\
& +  \left \{\frac{1}{3} \vec{\sigma}_1 \cdot \vec{\sigma}_2 q^2 \right \} 
\end{eqnarray}
%
%
The scalar part of the interaction we are interested in becomes
\begin{eqnarray}
\nonumber V_\pi^{scalar}(q) & = & \frac{g_\pi^2}{3}(\vec{\tau}_1 \cdot \vec{\tau}_2) (\vec{\sigma}_1 \cdot \vec{\sigma}_2) \frac{q^2}{m_\pi^2 + q^2} \\
 & & \hspace{-10mm} = \frac{g_\pi^2}{3}(\vec{\tau}_1 \cdot \vec{\tau}_2) (\vec{\sigma}_1 \cdot \vec{\sigma}_2) \left (1 - \frac{m_\pi^2}{m_\pi^2 + q^2} \right )
\end{eqnarray}

Fourier transformation of this and letting $\rr = |\mathbf{r}_2 - \mathbf{r}_1 | $ will give us the coordinate space representation
\begin{eqnarray}
\nonumber V_\pi^{scalar}(\rr)  & =  & \frac{g_\pi^2}{3} \left (\vec{\tau}_1 \cdot \vec{\tau}_2 \right ) (\vec{\sigma}_1 \cdot \vec{\sigma}_2) \times \\
& &  \left [ -4\pi \delta (\rr) + m_\pi^2 \frac{1}{\rr} e^{-m_\pi \rr} \right ]
\end{eqnarray}
As $m_q \rightarrow 0$, $g_\pi$ is finite and so we can neglect its dependence
 on $m_q/\Lambda_{QCD}$.
Now the strong force short range repulsion (implying proton and
 neutron hard cores) needs to be taken into account. It means that
 $\rr \neq 0$ and hence $\delta (\rr) = 0$. Nucleon core repulsion
is  incorporated into the
 interaction using the factor $f(\rr)$ which is presented e.g. in
 Ref. \cite{ref:nuc} 

\[\tilde{ V }(\rr) = {[f(\rr)]}^2 V(\rr)\]
where
\[  f(\rr)=1-{e^{- 1.1\,{\rho}^{2}}}\left (1- 0.68\,{\rho}^{2}\right ) \mbox{ \; , \;} \rho = \frac{\rr}{\mbox{fm}}.\]
Clearly, this factor restricts nucleon interaction at very short ranges, with $f(\rr = 0) = 0$, yet its effect is minimal at larger distances since $f(\rr) \approx 1$ for $\rr > 1 $ fm. It is this factor which results in a non-zero dependence of b on $m_q/\Lambda_{QCD}$.

Therefore we have the effective spin-spin interaction 
\begin{equation}
V_{ss} = V_o (\rr)(\mathbf{S}_1 \cdot \mathbf{S}_2) \times \mbox{\it{constant}}
\end{equation}
where
\begin{equation}
V_o (\rr) = m_\pi^2 \, e^{-m_\pi \rr} \frac{[f(\rr)]^2}{\rr} 
\end{equation}
We can obtain the  short range limit of this effective  interaction (which takes into account the finite size of the nucleons, that is the short-range repulsion). For clarity in our equations, we define $B$, $S$ and $S_1$. First let
\[ V_o (\rr) \approx B \, \delta (\rr)\]
so that we have 
\[  B  \:  = \: B \int \! \delta (\rr){d^3 \rr} \: = \: \int \! V_o (\rr){d^3 \rr} \]
Also let 
\begin{eqnarray}
\nonumber S & = & \frac{1}{m_\pi^2} \int \! V_o (\rr){d^3 \rr}  \\
 & = & 4\pi \int _{0}^{\infty}\!{e^{-m_\pi \rr}}\left [f(\rr)\right ]^{2}{\rr}{d\rr}
\end{eqnarray}
and
\[S_1  =  4\pi \; m_\pi \int _{0}^{\infty }\!{e^{-m_\pi \rr}}\left [f(\rr)\right ]^{2} [{\rr}]^2{d\rr} \]
so that 
\[ \frac{\partial S}{\partial m_\pi}  =  - \frac{S_1}{m_\pi}. \]
Thus
\[ \frac{\delta S}{S} =  -\frac{\delta m_\pi}{m_\pi}\frac{S_1}{S}. \]
From these definitions, we have
\[ B = m_\pi^2 \cdot S \]
and so
\[ \frac{\delta B}{B}\: = \: 2\frac{\delta m_\pi}{m_\pi} \: + \: \frac{\delta S}{S} \: = \: 2\frac{\delta m_\pi}{m_\pi} \: - \frac{\delta m_\pi}{m_\pi} \: \frac{S_1}{S}\]
Recalling that the $\pi-$meson contributes only 1/3 to the spin-spin
 interaction, we have:
\[ b  \sim  \left ( \frac{ \langle o | V_{ss} | k \rangle }{E_o - E_k} \right )^2 \: \sim \frac{1}{3}B^2 \mbox{ \; \: and  with \; } \frac{\delta m_{\pi}}{m_{\pi}} = \frac{1}{2} \frac{\delta m_q}{m_q }\]
we see
\begin{eqnarray}
\frac{\delta b}{b}= \frac{2}{3}\left ( 2 - \frac{S_1}{S}\right )
\frac{\delta m_{\pi}}{m_{\pi}} =\frac{1}{3}\left ( 2 - \frac{S_1}{S} \right )\frac{\delta m_q}{m_q}
\end{eqnarray}

\noindent
The integrals for $S$ and $S_1$ can be evaluated using $m_\pi = m_{\pi^o} = 135 \mbox{MeV} = 0.68 \mbox{\, fm}^{-1}$ to give
%
\[ \frac{S_1}{S} = 2.17. \] 
This gives a small number for $\frac{S_1}{S}-2 = 0.17$, therefore
the result may seem to be unstable. It is useful to clarify
this point using a simpler analytical model for the repulsive core
with  $f^2=1-\exp{(-kr)}$ which gives
\begin{eqnarray}\label{f}
\frac{S_1}{S} = 2 \frac{1-R^3}{1-R^2} \,\\
R=\frac{m_{\pi}}{k+m_{\pi}} \, .
\end{eqnarray} 
Any value of   $k>m_{\pi}$ 
gives $R<0.5$ and so gives a small difference for $\frac{S_1}{S}-2$.
Therefore, the small value of this difference
  does not indicate any strong instability.
If we take $k=1.1$ fm$^{-1}$ (the same value of the core radius
 which we used in the more sophisticated model for $f$ described above),
we obtain  $\frac{S_1}{S} = 2.2 $, i.e. practically the same result as above.
Thus, the result does not have any strong model dependence.
  
Using $\frac{S_1}{S} = 2.17$ we obtain
 the following $\pi$-meson contribution  to the variation of  $b$:
\begin{equation}
\left ( \frac{\delta b}{b} \right )_{\pi} =
 - 0.057 \; \frac{\delta m_q}{m_q}. 
\end{equation}
Similarly, for the $\rho$-meson we obtained $S1/S = 3.77$ and 
\begin{equation}
\left ( \frac{\delta b}{b} \right )_{\rho} =
-2.4\frac{\delta m_{\rho}}{m_{\rho}}
 = - 0.05 \; \frac{\delta m_q}{m_q}. 
\end{equation}
The final estimate is
\begin{equation}
\left ( \frac{\delta b}{b} \right )_{total}=- 0.11 \; \frac{\delta m_q}{m_q}.
\end{equation}
Note that if we assume that the spin-spin interaction is completely dominated
by the $\pi$-meson exchange the result (-0.17) would not be very different.

Returning to the magnetic moment, we have for the case of a valence proton:
\begin{eqnarray}
\nonumber \mu = g_n \, b \, \sz + (g_p - 1) (1-b) \sz + \jzp
\end{eqnarray}
\begin{eqnarray}
\nonumber \frac{\delta \mu}{\mu} =  \frac{\delta g_n}{g_n} K_n + \frac{\delta g_p}{g_p} K_p + \frac{\delta b}{b} K_{b_p}.
\end{eqnarray}
\begin{eqnarray}
K_{b_p}=  \frac{(g_n - g_p + 1) \szn}{\mu} 
\end{eqnarray}
Similarly, for the case of a valence neutron:
\begin{eqnarray}
\nonumber \mu = g_n (1-b) \sz + (g_p - 1) \, b \, \sz .
\end{eqnarray}
\begin{eqnarray}
\nonumber \frac{\delta \mu}{\mu} =  \frac{\delta g_n}{g_n} K_n + \frac{\delta g_p}{g_p} K_p + \frac{\delta b}{b} K_{b_n}.
\end{eqnarray}
\begin{eqnarray}
K_{b_n}=  \frac{(g_p - g_n - 1) \szp}{\mu} 
\end{eqnarray}
The dependence of $\delta \mu / \mu$ on the spin-spin interaction
 (via the  $K_b$ term) can now be seen to be quite significant.
 It depends on three values.
The first is the common factor $(g_p - g_n - 1) = 8.41$, which is large.
Next, it depends on the value of the effective spin of the
 \textit{non}-valence nucleons, which indicates the extent of the
 spin-spin interaction. It is most significant when the experimental value
 for $\mu$ deviates greatly from the valence model value.
Third, it depends on the nuclear magnetic moment and so is further enhanced
 when dealing with nuclei with small magnetic moments
 (e.g. $^{111}\mbox{Cd}$ has $\mu=-0.5949$ whereas $^{133}$Cs has
 $\mu = 2.582$).

We now have a modified version of equation (\ref{eq:mu5}) which includes
 the term $-0.11 K_{b}$ to account for the variation of the spin-spin
 interaction itself:
\begin{eqnarray} \label{eqn:kappa}
\nonumber \frac{\delta \mu}{\mu} = \kappa \frac{\delta (m_q/\Lambda_{QCD})}{(m_q/\Lambda_{QCD})}\\
 \kappa=-0.12K_n - 0.10K_p - 0.11K_b
\end{eqnarray}
\vspace{4mm}
\begin{table*}
\caption{Comparison of results for $\kappa$ (see equation (\ref{eqn:kappa})) for the three methods used in various nuclei}
\label{tab:c}
\vspace{2mm}
\centering

\begin{tabular}{@{\extracolsep{5mm}}c|ccccccc}
\hline
\hline \\[-2mm]
Atom  &  $^{87}$Rb & $^{111}$Cd & $^{129}$Xe & $^{133}$Cs & $^{139}$La &$^{171}$Yb & $^{199}$Hg \\
\\[-3mm] \hline \\[-2.5mm] 
Method A$^*$ & -0.074 &  -0.117 &  -0.117 &  0.127  & 0.127  & -0.117 &  -0.117 \\
\\[-3mm] \hline \\[-2.5mm] 
Method B$^*$ & -0.056 &  -0.111 &  -0.113 &  0.044 & 0.032 &-0.116 &  -0.116 \\
\\[-3mm] \hline \\[-2.5mm] 
Method C$^*$ & -0.016 &  \hspace{0.1mm} 0.125 &  \hspace{0.2mm} 0.042 &   0.009 & -0.008 & -0.085 &  -0.088 \\
\hline 

\end{tabular}

\; \footnotesize{$^*$ Method A considered only the valence nucleon, Method B includes the non-valence nucleons, \\ Method C further includes the effect of quark mass on the spin-spin interaction.}
\end{table*}

For ${}^{133}$Cs, we use the values obtained earlier to get
\[ K_{b_p} = \frac{(g_n - g_p + 1) \szn}{\mu}  = 0.335 \]

giving
\[ \frac{\delta\mu}{\mu} = 0.009 \frac{\delta (m_q/\Lambda_{QCD})}{(m_q/\Lambda_{QCD})} \]

%
%
%
%
%
%
%
%
%
%
%
%

All calculations for  ${}^{139}$La, ${}^{87}$Rb, ${}^{199}$Hg, ${}^{171}$Yb,
${}^{111}$Cd and ${}^{129}$Xe are similar to the method used for  ${}^{133}$Cs.
The results are presented in Table \ref{tab:c}, which 
summarizes the three methods used. 
Method A was the first method discussed
and used the theoretical nuclear magnetic moment of just the valence nucleon.
Method B included the contribution from non-valence nucleons to the nuclear
magnetic moment. Method C further included the effect of a variation in 
quark mass on the spin-spin interaction itself. It shows the significance
of the spin-spin interaction on how $\mu$ varies with quark masses, 
with sign reversal for some nuclei. 

\section*{Limits on the variation of the fine structure constant $\alpha$ and $(m_q/\Lambda_{QCD})$ using recent atomic clock experiments}

We can now estimate the time dependence of the ratio of the hyperfine
 transition frequencies to variations in $m_q/\Lambda_{QCD}$. The results for
 each atom M can be presented using the parameter $V$ as defined earlier with
 equation (\ref{eqn:V def}), with the values of $K_{rel}$ and $\kappa$ for the
 atoms considered here summarized in Table \ref{tab:e}.
\[ \frac{\delta V(M)}{V(M)} = \frac{\delta (A/E)}{A/E} =  \alpha^{2+K_{rel M}}
 \left(\frac{m_q}{\Lambda_{QCD}}\right) ^{\kappa_{_M}}  \frac{m_e}{m_p} . \]

For two atoms, $M_1$ and $M_2$, the dependence of the ratio of the frequencies $A(M_1)/A(M_2)$  can be presented as the ratio $X$ 
\begin{eqnarray} \label{eqn:X sum}
\nonumber X(M_1/M_2) & = & \frac{V(M_1)}{V(M_2)} \\ & = & \alpha^{K_{rel M_1}-K_{rel M_2}} \left(\frac{m_q}{\Lambda_{QCD}}\right) ^{\kappa_{_{M_1}} - \kappa_{_{M_2}}}
\end{eqnarray}

\begin{table*}
\caption{Summary of final results showing the relative sensitivity of the hyperfine relativistic factor to a variation in $\alpha$ (parameter $K_{rel}$) and the relative sensitivity of the nuclear magnetic moment to a variation in the quark mass/strong interaction scale $m_q/\Lambda_{QCD}$ (parameter $\kappa$). These values can be used in equation (\ref{eqn:X sum}).}
\label{tab:e}
\vspace{2mm}
\centering
\begin{tabular}{@{\extracolsep{3mm}}c|cccccccccc}
\hline
\hline \\[-2.5mm]
Atom   &  $_1^1$H &  $_1^2$H &  $_{2}^{3}$He    &  $_{37}^{87}$Rb  &  $_{48}^{111}$Cd &  $_{54}^{129}$Xe &  $_{55}^{133}$Cs & $_{57}^{139}$La & $_{70}^{171}$Yb &  $_{80}^{199}$Hg\\
\\[-3mm] \hline \\[-2.5mm]
$K_{rel}$   &    0    &   0   &    0    &   0.34     &   0.6   &  0.8  &   0.83    &   0.9  &   1.5   &  2.28\\
\\[-3mm] \hline \\[-2.5mm]
$\kappa$  & -0.100  & -0.064  &  -0.117  &  -0.016 & 0.125  & 0.042 & 0.009 & -0.008 & -0.085 &  -0.088\\
\hline
\end{tabular}
\end{table*}

For $A(^{87} \mbox{Rb})/A(^{133} \mbox{Cs})$, we have 
\begin{equation}
X(\mbox{Rb/Cs}) = \alpha^{-0.49} \left(\frac{m_q}{\Lambda_{QCD}}\right)
 ^{-0.025}
\end{equation}
and the result of measurements by \cite{Bize 2005} can be presented as a limit on the variation of X:
\[ \frac{1}{X(\mbox{Rb/Cs})} \frac{dX(\mbox{Rb/Cs})}{dt} = (-0.5 \pm 5.3)\times10^{-16}/\mbox{yr}. \]

For $A(^{133} \mbox{Cs})/A(^{1} \mbox{H})$, we have 
\begin{equation}
X(\mbox{Cs/H}) = \alpha^{0.83} \left(\frac{m_q}{\Lambda_{QCD}}\right) ^{0.109}
\end{equation}
and the result of the measurements in Ref.~\cite{Cs}  may be presented as
\begin{equation}\label{limitCsH}
 |\frac{1}{X(\mbox{Cs/H})}\frac{dX(\mbox{Cs/H})}{dt}|<
 5.5 \times 10^{-14}/yr \, . 
\end{equation}
For $A(^{171} \mbox{Yb}^+)/A(^{133} \mbox{Cs})$, we have 
\begin{equation}
X(\mbox{Yb$^+$/Cs}) = \alpha^{0.67} \left(\frac{m_q}{\Lambda_{QCD}}\right) ^{-0.093}
\end{equation}
and the result of measurements by \cite{Peik 2004} can be presented as a limit on the variation of X:
\[ \frac{1}{X(\mbox{Yb$^+$/Cs})} \frac{dX(\mbox{Yb$^+$/Cs})}{dt} = (2.8 \pm 2.9)\times10^{-14}/\mbox{yr}. \]

For $A(^{199} \mbox{Hg})/A(^{1} \mbox{H})$, we have 
\begin{equation}
X(\mbox{Hg/H}) = \alpha^{2.28} \left(\frac{m_q}{\Lambda_{QCD}}\right) ^{0.012}
\end{equation}
and the result of measurements by \cite{Prestage 1995} can be presented as a limit on the variation of X:
\[ \left | \frac{1}{X(\mbox{Hg/H})} \frac{dX(\mbox{Hg/H})}{dt} \right | < 8\times10^{-14}/\mbox{yr}. \]

The optical clock transition energy E(Hg) ($\lambda = 282$nm) in the Hg$^+$ ion can be presented in the form:
\begin{equation}
E(\mbox{Hg}) = const \times (\frac{m_e e^4}{\hbar^2})\, F_{rel}(Z\alpha)
\end{equation}
and calculations by Ref \cite{Dzuba 1999} gives
\begin{equation}
\frac{\delta E(\mbox{Hg})}{E(\mbox{Hg})} = -3.2 \frac{\delta \alpha}{\alpha}
\end{equation}
corresponding to $V(\mbox{Hg Opt})= \alpha^{-3.2}$. Variation of the ratio of the hyperfine splitting $A$(Cs) is given by 
\begin{equation}
V(\mbox{Cs})  =  \alpha^{2.83} \left(\frac{m_q}{\Lambda_{QCD}}\right) ^{0.009} \left(\frac{m_e}{m_p}\right) .
\end{equation}
 The relative variation of the electron to proton mass ratio can be described by \cite{Flambaum 2004}
\[ X(m_e/m_p) =   \left(\frac{m_q}{\Lambda_{QCD}}\right) ^{-0.037} \left(\frac{m_s}{\Lambda_{QCD}}\right)^{-0.011}  \frac{m_e}{\Lambda_{QCD}} .\]

giving
\begin{equation}
V(\mbox{Cs})  =  \alpha^{2.83} \left(\frac{m_q}{\Lambda_{QCD}}\right) ^{-0.039} \left(\frac{m_e}{\Lambda_{QCD}}\right)
\end{equation}

Variation of the ratio of the hyperfine splitting $A$(Cs) to this optical transition energy is given by:
\begin{eqnarray}
X(\mbox{Opt}) = \frac{V(\mbox{Cs})}{V(\mbox{Hg Opt})} = \alpha^{6}
 \left(\frac{m_q}{\Lambda_{QCD}}\right) ^{0.009} \left(\frac{m_e}{m_p} \right)\\ = \alpha^{6} \left(\frac{m_q}{\Lambda_{QCD}}\right) ^{-0.039} \left(\frac{m_e}{\Lambda_{QCD}} \right)
\end{eqnarray}
and the result of measurements by \cite{Bize 2003} can be presented as a limit on the variation of X:
\[ \left | \frac{1}{X(\mbox{Opt})} \frac{dX(\mbox{Opt})}{dt} \right | < 7 \times 10^{-15}/\mbox{yr}. \]

For the $1s-2s$ transition in hydrogen the relativistic corrections
 are negligible, i.e. $V(\mbox{H Opt})= \alpha^{0}$. 
Variation of the ratio of the hyperfine splitting $A$(Cs)
 to this optical transition energy is given by:
\begin{eqnarray}
X(\mbox{Opt}) = \frac{V(\mbox{Cs})}{V(\mbox{H Opt})} = \alpha^{2.83}
 \left(\frac{m_q}{\Lambda_{QCD}}\right) ^{0.009} \frac{m_e}{m_p} \\
 = \alpha^{2.83} \left(\frac{m_q}{\Lambda_{QCD}}\right) ^{-0.039} \left(\frac{m_e}{\Lambda_{QCD}} \right)
\end{eqnarray}
and the result of measurements by \cite{Fischer 2004} can be presented as a limit on the variation of X:
\[ \frac{1}{X(\mbox{Cs/H Opt})} \frac{dX(\mbox{Cs/H Opt})}{dt} = (3.2 \pm 6.3)\times10^{-15}/\mbox{yr}. \]

For the optical clock transition energy E(Yb) ($\lambda = 436$nm) in the Yb$^+$
ion calculations by Ref \cite{Dzuba 2003} gives
\begin{equation}
\frac{\delta E(\mbox{Yb})}{E(\mbox{Yb})} = 0.88 \frac{\delta \alpha}{\alpha}
\end{equation}
corresponding to $V(\mbox{Yb Opt})= \alpha^{0.88}$.
Variation of the ratio of the hyperfine splitting $A$(Cs) to this optical transition energy is given by:
\begin{eqnarray}
X(\mbox{Opt}) = \frac{V(\mbox{Cs})}{V(\mbox{Yb Opt})} = \alpha^{1.95}
 \left(\frac{m_q}{\Lambda_{QCD}}\right) ^{0.009} \frac{m_e}{m_p} \\
 = \alpha^{1.95} \left(\frac{m_q}{\Lambda_{QCD}}\right) ^{-0.039} \left(\frac{m_e}{\Lambda_{QCD}} \right)
\end{eqnarray}
and the result of measurements by \cite{Peik 2005} can be presented as a limit on the variation of X:
\[  \frac{1}{X(\mbox{Opt})} \frac{dX(\mbox{Opt})}{dt}=(1.2 \pm 4.4) \times 10^{-15}/\mbox{yr}. \]

Other combinations have been suggested as possible areas of research. For $A(^{129} \mbox{Xe})/A(^{3} \mbox{He})$, we have 
\begin{equation}
X(\mbox{Xe/He}) = \alpha^{0.8} \left(\frac{m_q}{\Lambda_{QCD}}\right) ^{0.159}
\end{equation}
while for $A(^{1} \mbox{H})/A(^{2} \mbox{H})$, we have 
\begin{equation}
X(^{1} \mbox{H} / ^{2}\mbox{H}) = \left(\frac{m_q}{\Lambda_{QCD}}\right) ^{-0.036}
\end{equation}

One can use Table \ref{tab:e} to predict which hyperfine transitions will be most sensitive to a variation in $\alpha$. The greatest effect will be seen for ratios between atoms with the greatest difference in values of $K_{rel}$ and $\kappa$, especially if relation (\ref{eq:alpha}) were correct. 
Clearly, it would be best to test ratios of elements with opposite signs for
 $\kappa$ so that the effects are more pronounced.

The effect of the spin-spin interaction is to reduce the sensitivity of Cs to
 a variation in quark mass and enhance the sensitivity of other nuclei
 such as  Cd. Because the spin-spin interaction is so strong for Cd,
 with its magnetic moment of -0.59 being quite different to the valence
 model value of -1.9, it may be quite sensitive to a variation in quark
 mass. In Ref. \cite{Karshenboim 2000}, the importance of Cd was motivated
by its small magnetic moment. This could enhance its sensitivity
to a variation of the fundamental constants. We have not actually obtained
any enhancement, and the absolute value of $|\kappa|$ in Cd is comparable to the
valence model value.
 However,  due to the spin-spin interaction
it has an opposite sign relative to some other nuclei with  large $|\kappa|$.
 For example, consider $A$(Cd)/$A$(H) and $A$(Cd)/$A$(He), with each ratio involving
 opposite signs for $\kappa$ :
\begin{equation}
X(\mbox{Cd/H}) = \alpha^{0.6} \left(\frac{m_q}{\Lambda_{QCD}}\right) ^{0.23}
\end{equation}
\begin{equation}
X(\mbox{Cd/He}) = \alpha^{0.6} \left(\frac{m_q}{\Lambda_{QCD}}\right) ^{0.24}
\end{equation}
Note that if relation (\ref{eq:alpha}) were correct, the variation of $X$
may be dominated by $m_q/\Lambda_{QCD}$: for $A$(Cd)/$A$(H), $X$(Cd/H)
 $\propto \alpha^{9}$.

\section{Conclusions}
The results of this work are presented in the previous section.
 Table \ref{tab:e} provides one with the numbers needed for the 
interpretation of the measurements. Below, we would like to formulate
a few conclusions which correct some misconceptions in the existing
literature and may help to plan future experiments and calculations.

$\bullet$ There is no such thing as a ``model-independent
interpretation of measurements'' if one uses the valence model
(Schmidt) values of the nuclear magnetic moments.
The valence model cannot even guarantee the order of magnitude and
sign of the effect.
  
   The situation may be improved by presenting the nuclear magnetic
moment as a linear combination of the neutron, proton and orbital magnetic
 moments. However, even this method does not guarantee high accuracy
since the expansion coefficients in this linear combination depend
on the ratio $m_q/\Lambda_{QCD}$. A consistent interpretation
of the measurements requires the calculation of the dependence
of nuclear magnetic moment on this parameter.

$\bullet$ A small value for the nuclear magnetic moment does not
guarantee an enhancement of the sensitivity to $m_q/\Lambda_{QCD}$.
However, a large deviation from the valence model value should 
increase the error in the calculated sensitivity.

$\bullet$ The dependence on $m_q/\Lambda_{QCD}$ of the nuclear magnetic moments
 of  $_{37}^{87}$Rb,   $_{55}^{133}$Cs, $_{57}^{139}$La
  and $_{54}^{129}$Xe is strongly
 suppressed by the many-body corrections - see Table \ref{tab:c}.
 One cannot guarantee high accuracy
of calculation in this situation. However, it is probably not important
since the suppression means that the contribution of these
magnetic moments to the final
effect of the variation is small. The effect will be dominated
by the variation of $\alpha$, $m_e/m_p$ or another magnetic moment.
 For $_{48}^{111}$Cd (where the magnetic moment
is small) the effect is not suppressed
 but it is of opposite sign to the valence model value.
For  $_{70}^{171}$Yb and  $_{80}^{199}$Hg the deviations
from the valence model are small. Naive calculations
give also reasonable results for  $_1^2$H and  $_{2}^{3}$He. 

\section{Acknowledgments}
    
    This work was supported by the Australian Research Council.
VVF is grateful to A. Brown, E. Eppelbaum, S.G. Karshenboim, E. Peik 
and V.G. Zelevinsky for useful discussions.

\end{document}